%% file: main.tex
\documentclass[10pt,twocolumn,letterpaper]{article}

\usepackage[pagenumbers]{cvpr} 
\usepackage[ruled,vlined]{algorithm2e}
\usepackage{pifont}


\definecolor{cvprblue}{rgb}{0.21,0.49,0.74}
\usepackage[pagebackref,breaklinks,colorlinks,allcolors=cvprblue]{hyperref}

\newcommand{\yuxin}[1]{\textcolor{black}{#1}}
\def\paperID{4035} 
\def\confName{CVPR}
\def\confYear{2025}

\title{VidMusician: Video-to-Music Generation with Semantic-Rhythmic Alignment via Hierarchical Visual Features
}

\author {
    Sifei Li\textsuperscript{\rm 1,\rm 2},
    Binxin Yang\textsuperscript{\rm 3},
    Chunji Yin\textsuperscript{\rm 3},
    Chong Sun\textsuperscript{\rm 3},
    Yuxin Zhang\textsuperscript{\rm 1,\rm 2},
    Weiming Dong\textsuperscript{\rm 1,\rm 2}\thanks{Corresponding author},
    Chen Li \textsuperscript{\rm 3}\\
    \textsuperscript{\rm 1}MAIS, Institute of Automation, Chinese Academy of Sciences \\
    \textsuperscript{\rm 2}School of Artificial Intelligence, University of Chinese Academy of Sciences\\
    \textsuperscript{\rm 3}WeChat, Tencent Inc.\\
}

\begin{document}
\maketitle
\input{sec/0_abstract}    
\input{sec/1_intro}

\input{sec/2_related_work}
\input{sec/4_Method}
\input{sec/3_Dataset}
\input{sec/5_Experiments}
\input{sec/6_Conclusion}
{
    \small
    \bibliographystyle{ieeenat_fullname}
    \bibliography{main}
}


\end{document}

%% file: sec/0_abstract.tex
\begin{abstract}
Video-to-music generation presents significant potential in video production, requiring the generated music to be both semantically and rhythmically aligned with the video.
Achieving this alignment demands advanced music generation capabilities, sophisticated video understanding, and an efficient mechanism to learn the correspondence between the two modalities.
In this paper, we propose VidMusician, a parameter-efficient video-to-music generation framework built upon text-to-music models. VidMusician leverages hierarchical visual features to ensure semantic and rhythmic alignment between video and music.
Specifically, our approach utilizes global visual features as semantic conditions and local visual features as rhythmic cues.
These features are integrated into the generative backbone via cross-attention and in-attention mechanisms, respectively.
Through a two-stage training process, we incrementally incorporate semantic and rhythmic features, utilizing zero initialization and identity initialization to maintain the inherent music-generative capabilities of the backbone.
Additionally, we construct a diverse video-music dataset, DVMSet, encompassing various scenarios, such as promo videos, commercials, and compilations.
Experiments demonstrate that VidMusician outperforms state-of-the-art methods across multiple evaluation metrics and exhibits robust performance on AI-generated videos. 
Samples are available at \url{https://youtu.be/EPOSXwtl1jw}.

\end{abstract}

%% file: sec/1_intro.tex
\section{Introduction}
\label{sec:intro}
If a video tells a story, then the background music sets its soul in motion.
Background music is extensively employed across domains such as advertising, film, animation, and short-form videos, significantly enhancing the content's appeal and viewer engagement. It serves as a key element in conveying the emotional tone of the video and providing an immersive experience for the audience.
Traditionally, video scoring relies on professional editors manually aligning video and audio, or on custom compositions from music producers, both of which are resource-intensive and inflexible.
In this work, we explore the extension of text-to-music models for generating music that aligns with video content, thereby reducing production costs and addressing potential copyright concerns.

The use of multimodal data for content generation has become a central theme in recent research. Several studies~\cite{yu2023long, li2024dance} investigate music generation synchronized with video, particularly for dance videos, by extracting rhythmic features from human motion. However, such human-centric approaches are not generalizable to the wide variety of videos found online.
Other works~\cite{di2021video, li2024diff} leverage symbolic music to generate background music for videos. Despite this, the generated music is often constrained to common instruments, lacking both melodic diversity and emotional depth, which creates a noticeable gap between these results and real music, limiting their applicability to diverse video content. Recent advances~\cite{liu2023m, tian2024vidmuse} focus on generating audio waveforms directly to produce background music for videos, but these methods either result in low-quality music or fail to achieve robust alignment between the video and the music.
Generating music that aligns with arbitrary videos is inherently challenging, as it requires both semantic coherence and rhythmic synchronization with diverse videos. Professional video editors often align visual variations with musical beats to enhance the overall audiovisual experience. However, prior works have not adequately addressed the issue of rhythmic alignment.

\input{Figures_tex/0_dataset}
To address the aforementioned challenges, we extend a text-to-music generation model into a video-to-music generation framework, aiming to produce music that aligns with various videos.
In this paper, we introduce VidMusician, a parameter-efficient framework that leverages hierarchical visual features to generate music aligned with both the semantics and rhythm of a video.
The visual encoder extracts global features, which capture semantic information, and patch-level local features, which reflect spatial visual variations.
Our approach employs a two-stage training process, incrementally incorporating global and local features into the network through cross-attention and in-attention mechanisms.
These features are progressively introduced as high-level semantic conditions and low-level rhythmic conditions, respectively, to guide music generation.
VidMusician is built on a pre-trained text-to-music model, requiring minimal parameter training to generate background music for videos.
To preserve the generative capabilities of the pre-trained model, we apply zero and identity initialization techniques when implementing the in-attention mechanism.

Another key obstacle is the lack of publicly available datasets that cover a wide range of scenarios.
To address this limitation, we collect a diverse video-music dataset, \textit{DVMSet}, which includes promo videos, commercials, compilations, animations, and various short-form videos.
Sample entries are illustrated in Figure~\ref{fig:dataset}.
Since online videos often contain voice tracks, non-musical elements, or noise, they are unsuitable for our task.
To remedy this, we develop a data collection process to curate high-quality video-music pairs.
This process involves removing vocal tracks and manually filtering the data.
As a result, we construct a diverse dataset comprising 3,839 video-music clips, which can fully validate the effectiveness and robustness of our method across different scenarios.

Our contributions can be summarized as follows:
\begin{itemize}[leftmargin=*]
\item We develop a multi-stage data collection process to obtain high-quality data from the internet and construct a diverse video-music dataset, \textit{DVMSet}, which covers a wide range of scenarios.

\item We propose VidMusician, a parameter-efficient framework that leverages hierarchical visual features to extend a text-to-music model into a video-to-music generation model. Our approach ensures both semantic and rhythmic alignment between the generated music and video content.
\item Our method introduces a two-stage training process that progressively incorporates high-level semantic control and low-level rhythmic control. To preserve the generative capabilities of the pre-trained model, we employ zero initialization and identity initialization when integrating rhythm conditions. Our approach achieves optimal results on both subjective and objective evaluation metrics.
\end{itemize}

%% file: Figures_tex/0_dataset.tex
\begin{figure*}[ht]
\centering
   \includegraphics[width=1\linewidth]{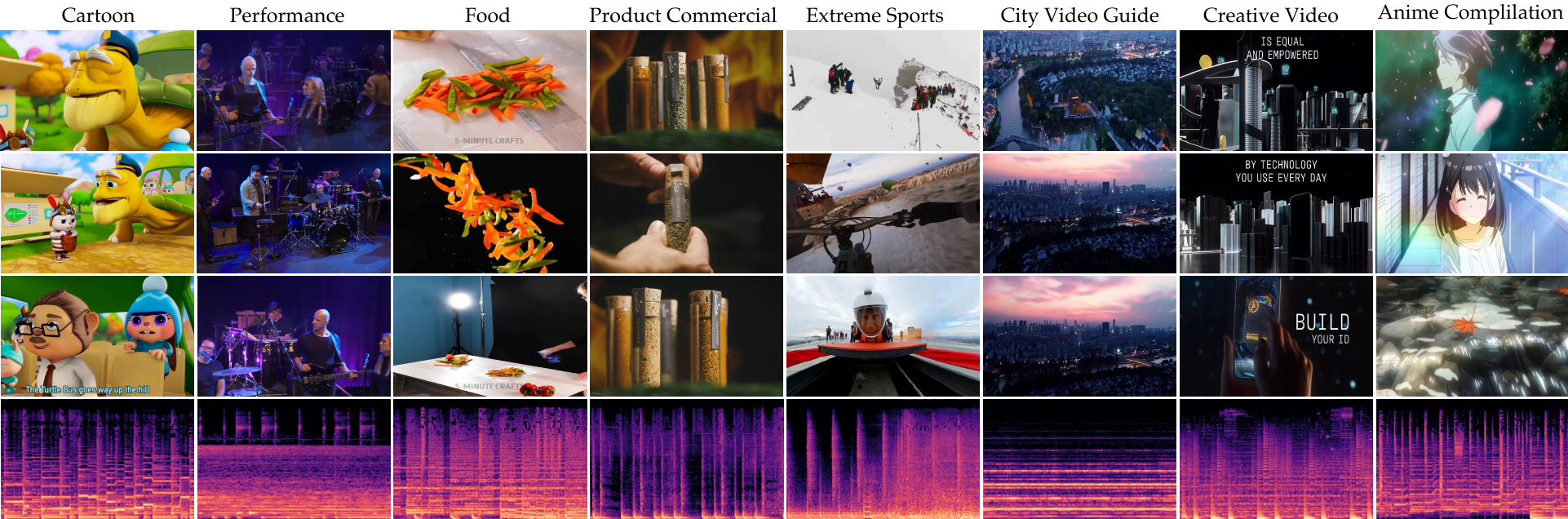}
   \caption{
   \yuxin{Real samples in \textit{DVMSet} which covers a wide range of visual scenarios and music types.}The first three rows display several video frames, while the last row shows the corresponding Mel-spectrogram segments of the music.}
\label{fig:dataset}
\end{figure*}

%% file: sec/2_related_work.tex
\section{Related Works}
\paragraph{Video-to-music Generation}
An increasing number of studies are exploring the generation of music that aligns with videos. Some works~\cite{zhu2022discrete, yu2023long, li2024dance, zhu2022quantized} focus on generating music for dance videos by extracting human motion information to control the rhythm of the generated music. However, these approaches are limited to videos with human motion, whereas our research addresses general video-to-music generation, applicable to a variety of scenarios.

Other studies~\cite{zhuo2023video} generate background music for videos based on symbolic music. The Controllable Music Transformer (CMT)~\cite{di2021video} uses rule-based rhythmic relationships to control timing, motion, beat, and music genre for background music generation. Video2Music~\cite{kang2023video2music} extracts semantic and motion features from videos, encodes them using a transformer encoder, and subsequently inputs them into a transformer decoder to autoregressively generate MIDI notes. 
Diff-BGM~\cite{li2024diff} uses visual features and textual descriptions of videos to guide a latent diffusion model in generating piano rolls with segment-aware cross-attention. However, these symbolic music-based methods face challenges in acquiring large-scale data and often produce music that significantly deviates from real music, exhibiting limited expressiveness and diversity.

More recent works~\cite{su2023v2meow} explore the direct generation of music waveforms from videos. 
M$^2$UGen~\cite{liu2023m}, a multimodal music understanding and generation model, uses a large language model as a bridge to facilitate video-to-music generation. VidMuse~\cite{tian2024vidmuse} constructs a large-scale dataset of 360K video-music pairs and employs a Long-Short-Term Visual Module to generate music aligned with the video. While these approaches focus on aligning video semantics with music, they lack fine-grained rhythm control. In contrast, our method uses an in-attention mechanism for rhythm control at the music token level.
VMAS~\cite{lin2024vmas} and MuVi~\cite{li2024muvi} explore rhythm alignment in video-to-music generation but require training generative backbones, unlike our parameter-efficient method. Consequently, there remains significant potential for exploration in video-to-music generation with semantic-rhythmic alignment.

\paragraph{Text-to-music generation}
Recent advancements in text-to-music generation have shown significant progress. While any-to-any generation frameworks such as CoDi~\cite{tang2024any} and NExT-GPT~\cite{wu2023next} are capable of producing content across multiple modalities, their effectiveness in text-to-music generation remains limited. Other approaches, such as Make-an-Audio~\cite{huang2023make}, AudioLDM~\cite{liu2023audioldm}, AudioLDM2~\cite{liu2023audioldm2}, and Tango~\cite{ghosal2023text}, utilize diffusion models to generate text-guided audio, including speech, sounds, and music. Despite their advancements in audio generation, these methods fall short of achieving the music generation capabilities of specialized text-to-music models.
Specialized models for text-to-music generation, such as diffusion-based models like Archisound~\cite{schneider2023mo}, Riffusion~\cite{Forsgren_Martiros_2022}, Moûsai~\cite{schneider2023mo}, Noise2Music~\cite{huang2023noise2music} and MusicLDM~\cite{chen2024musicldm}, succeed in generating high-quality music from text prompts. Transformer-based frameworks such as MusicLM~\cite{agostinelli2023musiclm}, MUSICGEN~\cite{copet2024simple}, and MuseCoco~\cite{lu2023musecoco} encode music as discrete tokens, enabling the creation of high-quality music with detailed textual control. Additionally, models such as QA-MDT~\cite{li2024quality}, Stable Audio2~\cite{evans2024long}, FluxMusic~\cite{fei2024flux}, and Seed-Music~\cite{bai2024seed} employ Diffusion Transformer (DiT) models to facilitate text-to-music generation.
In this work, we select the autoregressive text-to-music model MUSICGEN as our foundational model, considering its performance and accessibility. 

\paragraph{Controllable music generation}
Building on text-to-music generation models, many studies further explore methods for controllable music generation, which inspire our approach to leveraging video features to control music generation. Music ControlNet~\cite{wu2024music} adopts a strategy similar to ControlNet~\cite{zhang2023adding}, implementing various precise temporal controls on a pre-trained text-to-music diffusion model. \citet{li2024music} extends textual inversion~\cite{gal2022image} to time-varying textual inversion, employing a bias-reduced stylization technique for example-based music style transfer. CocoMulla~\cite{lin2023content} and AirGen~\cite{lin2024arrange} design parameter-efficient fine-tuning (PEFT) methods based on MUSICGEN~\cite{copet2024simple} to achieve content-based music generation and editing. Instruct-MusicGen~\cite{zhang2024instruct} utilizes a text fusion module and an audio fusion model for instruction-based music editing. MusiConGen~\cite{lan2024musicongen} employs an in-attention mechanism and designs an efficient fine-tuning strategy to control rhythm and chords in music. Our method is also a parameter-efficient framework, inspired by the controllable music generation methods~\cite{li2024music,lan2024musicongen}.

%% file: sec/4_Method.tex
\section {Method}
\input{Figures_tex/1_pipeline}

We propose a parameter-efficient video-to-music generation method that leverages hierarchical visual features to fully exploit the potential of visual cues in guiding both the semantic and rhythmic aspects of music. As illustrated in Figure~\ref{fig:pipeline}, our approach builds upon a pre-trained text-to-music model and incorporates a semantic conditioning module and a rhythm conditioning module to direct the music generation process. The semantic conditioning module employs an embedding manager and T5~\cite{raffel2020exploring} to encode global visual features, which are extracted at a sparse frame rate, into global semantic conditions.
The rhythm conditioning module leverages inter-frame similarity to aggregate local visual features extracted at a dense frame rate, thereby capturing spatial visual variations in the video. This module achieves low-level control through an in-attention mechanism. We employ a two-stage training process that incrementally integrates semantic and rhythmic conditions to help the model learn the correspondence between video and music. 

\subsection{Semantic Conditioning Module}
Video-to-music generation necessitates the alignment of both semantic and rhythmic dimensions. Semantic alignment pertains to the consistency between the music's style, emotion, and atmosphere with the video's semantics, which corresponds to the high-level features of the music.
Given that semantic changes within a video are often sparse and high-level musical features do not require fine-grained temporal information, we extract semantic features $V_S$ from the video at a frame rate of 1 fps. $V_S$ are derived from the CLS token of the Image Encoder $\mathcal{E}(\cdot)$ of CLIP~\cite{radford2021learning}:
\begin{equation}
V_S = \left[ s_i = \mathcal{E}(v_i)[0] \mid \forall i = 1, \ldots, N\right],
\end{equation}
where $v_i \in \mathbb{Z}^{3 \times H \times W}$ denotes a single frame of the video, $\quad s_i \in \mathbb{R}^{1\times D}$ denotes global features, and $N$ denotes the number of frames.

Pre-trained text-to-music models are proficient in controlling high-level musical features. Our objective is to harness this capability for video-to-music generation by refining it with video features.
We utilize an embedding manager $E$, composed of a few linear layers, to map visual features $V_S$ into the text embedding space.
Additionally, we finetune the T5 encoder $T$ using LoRA~\cite{hu2021lora} (Low-Rank Adaptation) to bridge the gap between visual and textual features.
The final semantic feature is defined as follows:
\begin{equation}
F_S = P(T(E(V_S))),
\end{equation}
where $P$ denotes the projector.
This approach seamlessly integrates video semantics into the generative model using a cross-attention mechanism, analogous to text conditions.
\input{Algorithm/0_Sim_process}
\input{Figures_tex/3_sim}
\input{Figures_tex/2_block_detail}

\subsection{Rhythm Conditioning Module}

Professionally edited videos often align music rhythm with visual changes, enhancing the audiovisual experience, similar to rhythm alignment in the video-to-music generation, which is relevant to low-level music features.
Unlike semantic alignment, rhythm alignment requires precise temporal control. Given that MUSICGEN~\cite{copet2024simple} performs autoregressive generation of music tokens, the ideal rhythm features should be at the level of the music token. 
We extract rhythm features $V_R$ at 25 fps, then raise to 50 fps to align with music tokens:
\begin{equation}
V_R = \left[ r_i = \mathcal{E}(v_i) \mid \forall i = 1, \ldots, N\right],
\end{equation}
where $r_i \in \mathbb{R}^{K\times D}$ denotes all token features.
We observed that, unlike the global features, which emphasize high-level semantics, local features extracted by CLIP, including image patch-level features, are sensitive to spatial variations within the image. As illustrated in Figure~\ref{fig:sim}, despite visual changes in the video, the inter-frame similarity of global features remains high, whereas local features exhibit more dynamic similarity, effectively capturing these changes. During the transition phase in Figure~\ref{fig:sim}(b), the similarity of local features drops significantly, while the similarity of global features shows a less pronounced decline. Consequently, we use the average inter-frame similarity of local features as the rhythm condition:
\begin{equation}
    sim = \frac{1}{K} \sum_{j=1}^{K}\cos(r_i^j, r_{i+1}^{j}).
\end{equation}
$sim$ is then processed according to Algorithm~\ref{alg:sim} to obtain $d$.
Although cross-attention offers a convenient method for incorporating conditions from other modalities, it is not well-suited for precise temporal control. This limitation arises from the inherent distribution of attention weights across multiple keys, which introduces ambiguity and hinders exact temporal alignment.
Additionally, the computational complexity of cross-attention increases quadratically with the sequence length ($O(L^2)$), making it computationally expensive to handle equally long rhythm and music token sequences.
To address these issues, we adopt the in-attention mechanism from MuseMorphose~\cite{wu2023musemorphose}, applying it similarly to MusiConGen~\cite{lan2024musicongen} by integrating it into the first self-attention layer of the initial three-quarters of the Transformer blocks, as shown in Figure~\ref{fig:block}. However, unlike MusiConGen~\cite{lan2024musicongen}, which fine-tunes the parameters of the first self-attention layers within each block, we fix all parameters of the generative backbone. We then utilize linear layers to differentiate the rhythm conditions across different blocks. The rhythmic condition of the $j$-th block can be defined as follows:
\begin{equation}
F_{Rj} = L_j(L_{j-1}(\cdots L_1(O(d)) \cdots)), \quad j = 1, \ldots, \frac{3}{4}N,
\end{equation}
where $N$ denotes the number of transformer blocks, $O$ represents a projector composed of a linear layer initialized with zeros, and $L_j$ are linear layers with weights initialized to the identity and biases to zero.
We define the output of $j$-th block $B$ as:
\begin{equation}
h_{j+1} = B(h_{j}+F_{Rj}),
\label{eq:T}
\end{equation}
where in the first training step, due to zero initialization and identity initialization, all \( F_{Rj} \) values are set to zero, making the generative model perform the same as the original pre-trained model.
Additionally, zero and identity initialization prevent the adverse effects of consecutive zero-initialized layers on gradient backpropagation. This approach efficiently incorporates rhythm features while preserving the generative capabilities of the pre-trained model.

\subsection{Two-stage Training}
To enhance the model's ability to decouple and comprehend different types of features, we employ a two-stage training approach to gradually introduce video semantic and rhythm features. Semantic features, which encompass the overall content and emotions of the video, are high-level and relatively stable. In contrast, rhythm features pertain to dynamic changes and are low-level and frequently fluctuating. In the first stage, we introduce semantic features to help the model understand the overall video content, training only the semantic conditioning module. In the second stage, we add rhythm features for comprehensive control, training both the semantic and rhythm conditioning modules. The use of zero initialization and identity initialization in the second stage allows us to seamlessly model rhythm without compromising the progress made in the first stage.

%% file: Figures_tex/1_pipeline.tex
\begin{figure*}[ht]
\centering
   \includegraphics[width=1\linewidth]{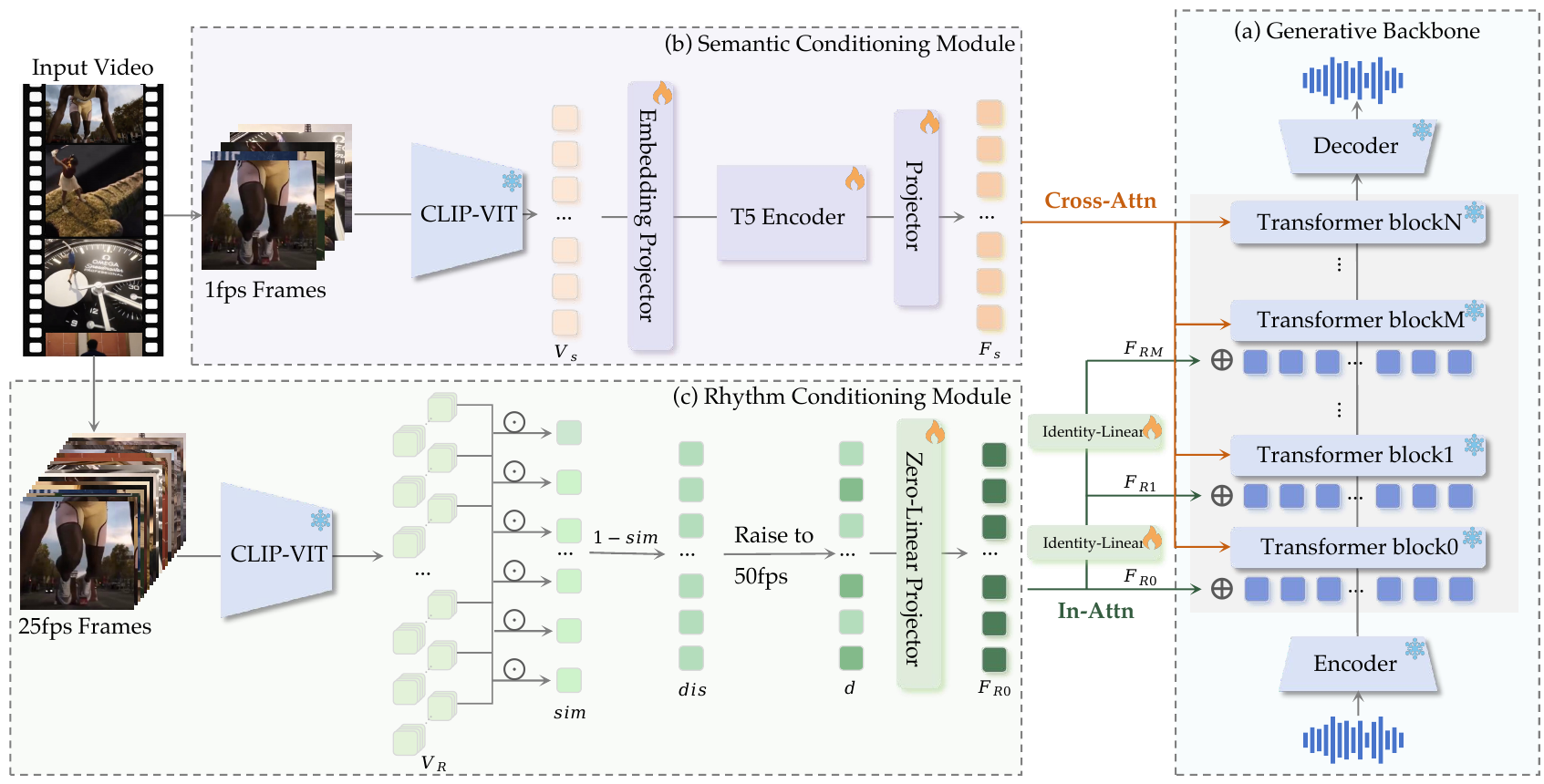}
   \caption{
\yuxin{
We employ an autoregressive model as the music \textbf{(a) Generative Backbone} and propose a method for semantic and rhythmic control via hierarchical visual features. 
The \textbf{(b) Semantic Conditioning Module} maps the CLIP global features into the text embedding space of the T5 model, which is fine-tuned using LoRA~\cite{hu2021lora}, and its output is incorporated into the generative backbone via a cross-attention mechanism.
The \textbf{(c) Rhythmic Conditioning Module} captures spatial variations by computing inter-frame cosine distances of CLIP local features, and its output is incorporated into the generative backbone via an in-attention mechanism. 
``$\odot$'' represents the cosine similarity calculation, 
$1 - sim$ represents the operation of subtracting the value of $sim$ from 1 to get $dis$, and ``$\oplus$'' represents element-wise addition. 
``Zero-Linear'' and ``Identity-Linear'' refer to linear layers with zero and identity initialization techniques, respectively. }
}
\label{fig:pipeline}
\end{figure*}

%% file: Algorithm/0_Sim_process.tex
\begin{algorithm}[!h]
    \caption{Compute Distance Sequence}
    \label{alg:sim}
    
    \KwIn{$sim$}  
    \KwOut{$d$}  
    
    $sim \leftarrow [1] + sim$ \tcp*{Keep the number of video frames}
    $dis \leftarrow [1 - s \text{ for } s \text{ in } sim]$ \tcp*{Get cosine distance}
    $d \leftarrow \sum([[i, 0] \text{ for } i \text{ in } dis], [])$ \tcp*{Raise to 50 fps to align with music tokens}
    
    \Return $d$ 
\end{algorithm}
    
    

%% file: Figures_tex/3_sim.tex
\begin{figure*}[ht]
\centering
   \includegraphics[width=1\linewidth]{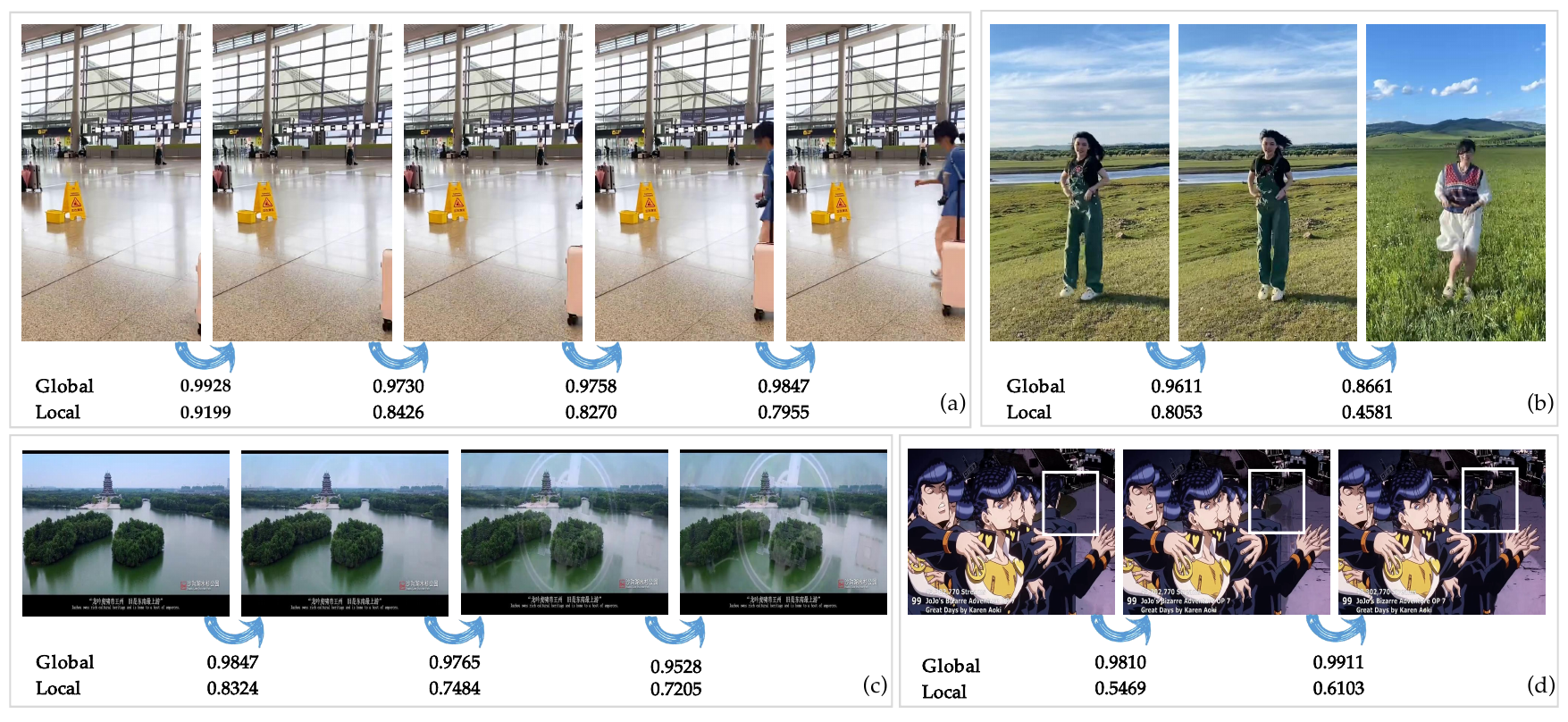}
   \caption{\textbf{Inter-frame similarity} is assessed using both global and local features, as indicated beneath each set of frames. The first row shows global similarity, while the second row depicts local similarity. In (a), (c), and (d), despite variations in the frames, the global similarity remains close to 1, whereas the local similarity varies with the degree of change. In (b), during transitions, both similarities decrease, but the local similarity exhibits a more pronounced decline. In (d), white boxes highlight the primary areas of change within the video frames. }
\label{fig:sim}
\end{figure*}

%% file: Figures_tex/2_block_detail.tex
\begin{figure}[ht]
\centering
   \includegraphics[width=0.7\linewidth]{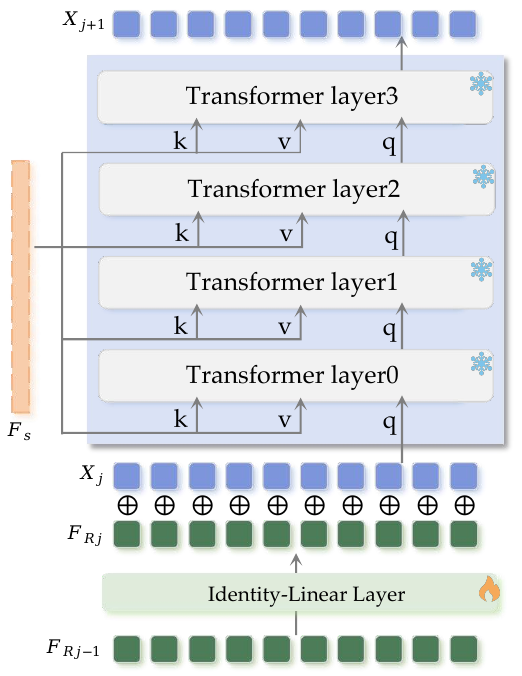}
   \caption{
   \yuxin{
    \textbf{Details of the Transformer block}. Each block consists of four Transformer layers. Cross-attention, which incorporates semantic condition, is applied at every layer, whereas in-attention, which integrates rhythm condition, is applied only at the first layer of each block.}}
\label{fig:block}
\end{figure}

%% file: sec/3_Dataset.tex
\section {Dataset}

Existing open-source video-music datasets~\cite{li2021ai, zhu2022quantized, hong2017content, zhuo2023video, li2024diff, liu2023m, li2018creating} are limited in diversity, predominantly focusing on specific content types such as dance videos, music performances, and music videos. To overcome these limitations, we construct a diverse video-music dataset, \textit{DVMSet}, by sourcing data from public media platforms. Through meticulous processing and filtering, we curate 3,839 high-quality music-video pairs. Despite the abundance of music videos online, their quality varies significantly, with many featuring non-musical audio. To ensure the inclusion of high-quality music videos across various categories, we craft specific queries such as ``food commercial'', ``opening title sequence'', and ``city video guide''. These queries primarily aim to collect official or professionally edited music videos, thereby ensuring superior data quality.

We observe that the majority of the collected videos contain vocal tracks. To encompass a diverse range of scenarios, we do not restrict our dataset to instrumental music. Instead, we employ Demucs~\cite{rouard2023hybrid} to remove vocals and perform volume normalization for consistency. During data collection based on queries, the inclusion of noisy data or videos with minimal music content is inevitable. Current audio source separation models may also leave harsh residual noise after vocal removal. To ensure data quality, we first segment the data into 60-second clips, and then human experts manually filter and exclude low-quality samples based on criteria such as more than 5 seconds of non-music content, persistent noise, static images, and mismatched background music. Ultimately, we obtain 3,839 high-quality clips from 2,043 videos, out of a total of 11,126 clips. These are divided into 3,571 clips for training and 268 clips for testing, both containing samples from 47 queries. When splitting the data, we base the split on the original videos to ensure that clips from the same original video do not appear in both the training and testing sets. Figure~\ref{fig:dataset} illustrates some examples from the dataset, and we will release the URLs of the videos in the dataset.

%% file: sec/5_Experiments.tex
\section{Experienments}
\input{Tables/Contrast}

\subsection{Implementation Details}
In our experiments, we freeze the parameters of the music generation backbone. During the first stage, we train the embedding manager and projector of the semantic conditioning module and fine-tune the T5 model using LoRA~\cite{hu2021lora}. In the second stage, we extend the training by incorporating the rhythm conditioning module, initializing its linear layers with zero and identity initialization. Our approach achieves video-to-music generation with only 24.79M trainable parameters. We employ learning rates of 1e-4 and 1e-5 for the first and second stages, respectively. The AdamW optimizer is used with $\beta_1 = 0.9$, $\beta_2 = 0.95$, and a weight decay of 0.1. A warm-up learning rate is applied to all training layers during the initial 6000 iterations. The maximum duration of music generation is set to 30 seconds, consistent with MUSICGEN~\cite{copet2024simple}. We train for 120 epochs in each stage with a batch size of 16 on four NVIDIA Tesla V100-SXM2 GPUs.

\subsection{Evaluation Metrics}
To quantitatively evaluate our method, we employ a comprehensive set of metrics based on existing standards and further analysis of the task.

\paragraph{Audio Quality}
We utilize the Frechet Distance (FD) and Kullback-Leibler Divergence (KL) metrics to assess audio quality, both implemented using PANNs~\cite{kong2020panns}. FD measures the distance between the distributions of features from generated and real music, while KL quantifies the divergence between the class outputs of generated and real music.

\paragraph{Semantic Alignment}
We use the ImageBind~\cite{girdhar2023imagebind} and CLAP~\cite{wu2023large} scores for semantic alignment evaluation. The ImageBind score is calculated by computing the average cosine similarity of video and music features extracted from ImageBind. Similarly, the CLAP score is obtained by computing the average cosine similarity of real and generated music features.

\paragraph{Rhythm Alignment}
Currently, there is no dedicated metric for evaluating rhythm alignment between music and video. We propose a new metric that measures rhythm alignment by utilizing SceneDetect~\cite{Castellano_PySceneDetect} to identify frames with abrupt visual changes and detecting musical beats similar to the Beats Hit Score (BHS) in~\cite{zhu2022quantized}. Since abrupt visual changes in videos are typically sparse, and musical beats occur more frequently to maintain continuity, musical beats do not necessarily align with the visual changes. We calculate the recall rate of scene change frames at beats, allowing a tolerance of 0.1s. Furthermore, we exclude videos featuring stable visuals or soft background music, such as landscapes, which do not impose explicit requirements on musical rhythm, focusing instead on professionally edited videos.

\subsection{Quantitative Evaluation}
We compare our method with five state-of-the-art approaches: MIDI-based methods (CMT~\cite{di2021video}, Video2Music~\cite{kang2023video2music}, and Diff-BGM~\cite{li2024diff}) and waveform-based methods (M$^2$UGen~\cite{liu2023m} and VidMuse~\cite{tian2024vidmuse}). 
We evaluate MIDI-based methods using the models provided by the authors. For waveform-based methods, we conduct evaluations using both the authors' models and models trained on our dataset. The comparative results of our method with others are shown in Table~\ref{tab:contrast}. Our method achieves optimal performance across all metrics. Music generated by MIDI-based methods exhibits noticeably poorer audio quality, highlighting the advantage of directly generating audio over MIDI notes. Due to limited musical expressiveness, these methods perform poorly in semantic alignment. Video2Music, which uses SceneDetect~\cite{Castellano_PySceneDetect} for rhythm control, shows suboptimal performance in rhythm alignment. M$^2$UGen, when trained on our dataset, improves in ImageBind score and KL metrics but performs worse on other metrics. The original VidMuse model outperforms the model trained on our dataset, indicating its reliance on larger datasets for effective modeling. In contrast, our method leverages a pre-trained text-to-music model, achieving diverse video-to-music generation despite using a dataset much smaller than 360k. Our method has only 24.79M training parameters, substantially less than the 1,880.14M training parameters of VidMuse.

\paragraph{User study} We conduct a subjective evaluation of our method against other approaches.
We randomly sample 175 pairs of results, each containing one of our results and one result from another method for the same video. Participants are instructed to evaluate the music of each pair in terms of semantic alignment, rhythm alignment, and overall quality, selecting the better result (excluding rhythm comparisons for 35 pairs of videos, such as landscapes, which feature stable visuals and do not impose explicit rhythmic requirements). In total, we collect votes from 46 participants for these 175 pairs of results. The percentage of votes for each approach is shown in Table ~\ref{tab:contrast}. The row corresponding to our method shows the average percentage of votes compared to the other methods. Our method receives a higher percentage of votes across all three metrics, demonstrating its superiority in generating background music for videos.

\input{Tables/Ablation}

\subsection{Qualitative Evaluation}
The samples encompass a variety of scenarios, including cartoons, promos, commercials, compilations, dance performances, landscapes, and other creative videos, showcasing our model's adaptability to diverse scenes. Furthermore, it demonstrates superior rhythm alignment in professionally edited videos. Additionally, we validate the effectiveness of our method in generating background music for AI-generated videos. The successful application of AI-generated videos underscores the generalizability of our approach and highlights its potential for real-world applications.

\subsection{Ablation Study}
To demonstrate the effectiveness of each component of our method, we conduct comprehensive ablation experiments, as shown in Table~\ref{tab:ablation}.

\paragraph{From pre-trained T5}
We demonstrate the effectiveness of utilizing the high-level control capabilities of the pre-trained text-to-music model.
As indicated in the first two rows of Table~\ref{tab:ablation}, training the T5~\cite{raffel2020exploring} model from scratch results in a substantial drop in semantic alignment. This finding suggests that mapping visual features to the text embedding space of the pre-trained T5 model significantly enhances semantic control based on video features.

\paragraph{Initialization techniques} The third row of Table~\ref{tab:ablation} shows that without initialization techniques (IT), all metrics drop significantly, indicating that standard linear layer initialization disrupts the pre-trained generation model.

\paragraph{Two-stage training} The fourth row of Table~\ref{tab:ablation} shows that omitting the two-stage (TS) training approach degrades performance across various metrics, highlighting the importance of gradually introducing semantic and rhythm control for better understanding video features in music generation.

\paragraph{Visual features for rhythm control} The fifth row of Table~\ref{tab:ablation} shows that using only global features for rhythm control leads to a decline in rhythm-related metrics, demonstrating that leveraging local features, which encode spatial information, is more effective for rhythm control.

\paragraph{Comparison with MusiConGen} The sixth and seventh rows of Table~\ref{tab:ablation} show that using the MusiConGen~\cite{lan2024musicongen} strategy, where the first self-attention layer of each block is trained for rhythm control, significantly degrades performance without initialization techniques. Even with initialization, most metrics perform worse, validating the superiority of our rhythm control method over MusiConGen.


%% file: Tables/Contrast.tex
\begin{table*}
   \caption{Quantitative evaluation and user study results in comparison with state-of-the-art methods. Note that the subjective metrics of our method show the average vote percentage compared to other methods. ``*'' represents training the model with our dataset. The best results are in highlighted \textbf{bold} and the second best ones are \underline{underlined} (same in the following tables). }
   \begin{center}
   \begin{tabular}{c||cc|cc|c|ccc}
   \hline

  {}&\multicolumn{5}{|c}{Objective Metrics} & \multicolumn{3}{|c}{Subjective Metrics} \\
    \cline{2-9}
     & FD $\downarrow$ & KL  $\downarrow$ & ImageBind $\uparrow $ &CLAP  $\uparrow$& Rhythm $\uparrow$  & Semantic $\uparrow$ & Rhythm $\uparrow$ & Overall $\uparrow$\\ 
   \hline
   CMT~\cite{di2021video}  & 61.45  & 1.673  &0.1103 & 0.3064& 0.1637  &  11.83$\%$  &12.93$\%$ & 11.91$\%$\\
   Diff-BGM~\cite{li2024diff}  & 105.5  & 2.243 & 0.1209& 0.1905 &  0.1257 & 15.57$\%$ &14.46$\%$  & 15.39$\%$\\
   Video2Music~\cite{kang2023video2music}& 101.3  & 2.147 & 0.0783  &0.2221&  \underline{0.2381} &15.48$\%$ & 25.22$\%$ &17.22$\%$\\
   M$^2$ugen~\cite{liu2023m}  & 41.79 & 1.900 &0.1291 &0.2968& 0.2288 & 19.65$\%$  &18.48$\%$  &20.17$\%$\\
   M$^2$ugen*  & 45.46  & 1.691 &0.1429 &0.2913& 0.1915  &13.48$\%$ & 12.28$\%$  &13.83$\%$\\
   VidMuse~\cite{tian2024vidmuse} &  \underline{31.43} & \underline{1.260}  & \underline{0.1960} & \underline{0.3853} & 0.1722 &\underline{33.30$\%$}  & \underline{31.74$\%$} & \underline{31.83$\%$}\\
  VidMuse* & 33.85 & 1.521 & 0.1610 & 0.2900 & 0.2305  & 19.74$\%$ &23.37$\%$  & 19.57$\%$\\
  \textbf{VidMusician(ours)}  & \textbf{27.46} & \textbf{1.112}  & \textbf{0.2158} & \textbf{0.4162} &\textbf{0.2516} & \textbf{81.56$\%$} & \textbf{80.22$\%$}  &\textbf{81.44$\%$}\\
   \hline
   \end{tabular}
   \end{center}
   \label{tab:contrast}
\end{table*}

%% file: Tables/Ablation.tex
\begin{table*}
   \caption{Ablation study of our method. ``From T5'' indicates the use of pre-trained model parameters from T5, ``RCM'' represents the Rhythm Control Module, ``IT'' represents Initialization Technique, ``TS'' denotes Two-Stage Training,``AT'' represents using all tokens of CLIP for rhythm control and ``RM'' refers to rhythm control similar to MusiConGen~\cite{lan2024musicongen}.}
   \begin{center}
   \begin{tabular}{c ccccc|cc|cc|c}
   \hline
   From T5 & RCM & IT & TS & AT & RM  & FD  $\downarrow$ & KL  $\downarrow$ & ImageBind Score $\uparrow $ &CLAP score $\uparrow$ & Rhythm $\uparrow$\\
   \hline
  \ding{55} &\ding{55} & \ding{55} &\ding{55} &\ding{55} &\ding{55}  & 36.62  & 1.564  & 0.1480 &0.2690& 0.1392 \\
   \ding{51} &\ding{55} & \ding{55} &\ding{55}&\ding{55} &\ding{55} &  30.19 & \underline{1.107}  &0.1872&0.3536&0.1644   \\
   \ding{51} &\ding{51} & \ding{55} &\ding{51} &\ding{51}&\ding{51}  &81.70 & 1.613  &0.0803&0.1567& 0.1283  \\
   \ding{51} &\ding{51} & \ding{51} &\ding{55} &\ding{51} &\ding{51}& 33.57 & 1.369 & 0.1735&0.3614&  \underline{0.2431}  \\
   \ding{51} &\ding{51} & \ding{51} &\ding{51} &\ding{55} &\ding{51}& 29.28 & 1.156 & \underline{0.1993} &\underline{0.4126}& 0.2346  \\
  \ding{51} &\ding{51} & \ding{55} &\ding{51} &\ding{51} &\ding{55}  & 96.13 & 2.184 &  0.1066& 0.0933  & 0.1109 \\
   \ding{51} &\ding{51} & \ding{51} &\ding{51} &\ding{51} &\ding{55} & \underline{29.23} & \textbf{1.106} & 0.1892& 0.3731  & 0.1498 \\

  \ding{51} &\ding{51} & \ding{51} &\ding{51} &\ding{51} &\ding{51} & \textbf{27.46} & 1.112  &\textbf{0.2158} & \textbf{0.4162}&\textbf{0.2516}   \\
   \hline
   \end{tabular}
   \end{center}
   \label{tab:ablation}
\end{table*}



%% file: sec/6_Conclusion.tex
\section{Conclusion}
In this work, we introduce VidMusician, a parameter-efficient framework for video-to-music generation that leverages hierarchical visual features, developed from a pre-trained text-to-music model. To achieve semantic and rhythmic alignment, we design control modules that utilize global and local features, incorporating semantic and rhythmic conditions into the generation backbone via cross-attention and in-attention mechanisms, respectively. Our two-stage training strategy and initialization techniques fully exploit the generative capabilities of the pre-trained model while guiding it to progressively learn the mechanisms of video background music generation. Additionally, we construct a diverse video-music dataset, \textit{DVMSet}, encompassing various scenarios. Experimental results demonstrate that VidMusician generates high-quality music semantically and rhythmically aligned with various videos, outperforming current state-of-the-art methods and exhibiting robust performance even on AI-generated videos.
